\documentclass{PoS}

\usepackage{wrapfig}
\title{Unfolding the True Atmospheric Neutrino Event Rate in the 1Gev - 1Tev Range Using IceCube/DeepCore}

\ShortTitle{Atmospheric Event Rate Unfolding}

\author{
The IceCube Collaboration\footnote{For collaboration list, see PoS(ICRC2019) 1177.}\\
{\itshape \href{http://icecube.wisc.edu/collaboration/authors/icrc19_icecube}{http://icecube.wisc.edu/collaboration/authors/icrc19\_icecube}}\\
E-mail: \email{jsandroos@icecube.wisc.edu}
}

\abstract{An unfolding measurement of the atmospheric neutrino flux at the South Pole is performed using the IceCube/DeepCore detector. The main results presented is the true neutrino interaction rate by unit volume of ice, as a function of energy or zenith angle. The method used is the D'Agostini iterative unfolding. While the detector response estimate is based on Monte Carlo simulation, the iterative approach will compensate for this inherent bias and draw the unfolded result closer to the unbiased estimator, as the number of iterations is optimized. This is done using an ensemble test with a blind re-smearing approach. Thus this measurement is minimally biased regarding atmospheric flux-, oscillation- and neutrino interaction models, allowing model builders to test their predictions against this measurement. As an aside, using the same data set and methodology, we also present an unfolding measurement directly to atmospheric neutrino flux \\

\vspace{4mm}
{\bfseries Corresponding authors:}
\speaker{Joakim Sandroos}$^{1}$\\
{$^{1}$ \itshape J. G. U. Mainz}\\

}

\FullConference{36th International Cosmic Ray Conference -ICRC2019-\\
		July 24th - August 1st, 2019\\
		Madison, WI, U.S.A.}

\begin{document}

\section{Introduction}
IceCube is a cubic-kilometer neutrino detector installed in the ice at the geographic South Pole \cite{Aartsen:2016nxy} between depths of 1450 m and 2450 m, completed in 2010. Reconstruction of the direction, energy and flavor of the neutrinos relies on the optical detection of Cherenkov radiation emitted by charged particles produced in the interactions of neutrinos in the surrounding ice or the nearby bedrock. The DeepCore subarray as defined in this analysis includes 8 densely instrumented strings optimized for low energies plus 12 adjacent standard strings. This densely instrumented region pushed the energy threshold of the detector down to below 10 GeV. IceCube/DeepCore is sensitive to neutrinos generated by cosmic ray interactions in the atmosphere. These atmospheric neutrinos are utilized in many experiments to measure fundamental physical properties, most famously the neutrino oscillation parameters. The precision of any such measurement is directly dependent on the prior knowledge of the atmospheric neutrino flux and its associated precision. Usually measurements of the atmospheric neutrino flux proceeds from model assumptions via forward folding to parameter testing against a measurement. It is then not straight forward to deconvolute a flux measurement from oscillation parameters. It is then desirable to have available a model agnostic measurement of the atmospheric neutrinos, which makes as few as possible assumptions about the model parameters of interest. This is achieved through unfolding, and we present the unfolded in-ice event rate by volume, for different types of neutrino interaction channels as function of zenith and energy.

\section{Method}
We utilize the Bayesian unfolding method as developed by D'Agostini and implemented in the RooUnfold software package \cite{Adye:2011}. We give here a brief overview of the method - more details can be found in \cite{DAgostini:2010}. The main goal of unfolding is to compensate for detector effects and efficiency in an effort to revert the measurement back to the truth of nature. The detector resolutions are described by a 'response matrix' correlating the true and reconstructed values of the measured quantities. This unfolding method starts from Bayes theorem, to estimate the probability of a measured event in bin j to originate in true bin i:
\begin{equation}
    P(i|j) = \frac{P(j|i) P(i)}{P(j)}
\end{equation}
, where \(P(j|i)\) is the response matrix, \(P(i)\) is the prior probability for a true event to fall in bin i, and \(P(j)\) is the total probability for an event to fall in bin j. \(P(i|j)\) is known as the unfolding matrix and the unfolded result \(U_{n}\) will be obtained by applying the unfolding matrix to the measured spectrum, n times:
\begin{equation}
    U_{0} = P(i|j)_{0} \times MF
\end{equation}
The index marks the iteration in question. For every iteration the prior probability is updated with the unfolded result, and a new unfolding matrix is calculated, leading to the next iteration. Usually the unfolded spectrum converges rather quickly on the truth in about 4-5 iterations, however more can be needed \cite{DAgostini:2010}. A few details are worth mentioning: First, the response matrix is based on our MC simulation and in the ideal case each iteration moves the unfolded spectrum away from the prior assumption and closer to the unbiased estimate. Typically at 1-2 iterations the unfolding will show a strong bias towards the simulation used for constructing the response matrix, while higher iterations show better consistency with the truth of the input sample. This suggests more iterations are better in general.  However in rare cases it is possible for the unfolding to diverge from the MC truth. Careful investigation of the unfolding behaviour with number of iterations is necessary, and the number of iterations in the analysis is a trade-off: The statistical uncertainty of the unfolded spectrum arises from the uncertainty on the measurement as well as the unfolding matrix. With the iterative updating of the prior term in eq. 2.1 the statistical uncertainty is re-introduced in the unfolding matrix calculation, leading to an increasing uncertainty in the unfolding with the number of iterations. Moreover statistical fluctuations in the unfolded sample, will be fed back into the unfolding matrix calculation potentially creating a situation in which these fluctuations will grow with the number of iterations. Choosing a good number of iterations is thus a trade off between bias and variance. 
Knowledge of the response matrix is obtained by applying the detector reconstruction algorithm to a suitably large Monte Carlo sample, recording the simulation truth and reconstructed values for each event. This can be done in multiple dimensions, giving different requirements on the size of the MC sample. In order to be in a regime where the statistical uncertainty of the response matrix is insignificant compared to the impact from systematic uncertainties, we need on the order of \(2-5 \cdot 10^{5}\) MC events, as the matrix has ~83.000 bins in total. The MC sample here used however contains about 2 million events, leaving us well above that limit. The reconstructed quantities are the energy and zenith angle of the incoming neutrino as well as the track length signature of the outgoing lepton. The outgoing lepton only exists in the case of a charged current interaction, but the reconstruction algorithm will always reconstruct both an initial cascade and a track. In case of neutral current interactions the reconstructed track will be short, and thus the track length can be correlated with the flavor and interaction type (PID) of the incoming neutrino. Our response matrix thus has three dimensions in both true- and reconstructed space: Energy, zenith angle and PID. However since the track length is virtually identical for many flavor and interaction types the PID dimension will cover only two configurations: \(\nu_{\mu}^{cc} + \bar{\nu}_{\mu}^{cc}\) covering the muon neutrino and muon anti-neutrino charged current interactions, and \(\nu_{rest}\) covering all other flavor and interaction combinations. 

\section{Systematic Error}

The systematic uncertainties represent the precision in our knowledge of the experiment. 16 systematic parameters have been included in this analysis, as listed in table \ref{table:systematics}. 12 of them are related to physics and 4 are related to direct detector effects. The nominal values of oscillation parameters are taken from the global fit in  The response matrix shape changes with the systematic parameters resulting in an uncertainty on the calculated unfolding matrix, which in turn propagates to the unfolded spectrum. The individual pulls, as well as cross correlations between systematic parameters have been investigated. Out of these 16 only the optical efficiency of the optical modules and the scattering and absorption in the ice have a significant impact on the analysis. In order to estimate the final error due to systematics 1400 random trial sets have been generated, where for each set, each parameter was set to a randomized value based on a gaussian prior. The unfolding analysis was then run for each trial set. This procedure allows for the construction of an error band based on the coverage percentiles of the trials. Since the systematic impact is dominated by 3 systematic parameters, 1400 trials is sufficient to cover the parameter space adequately. Individual pulls range from 0.01\% to about 10\% at the \(\pm 1 \sigma\) confidence level. Depending on the shape and magnitude of the systematic variations, the iterative unfolding can pick up on these variations and make them grow with the number of iterations. This effect is similar to the statistics-only case, but seen only in select cases when the input measurement deviates several sigma from the MC.

\section{Closure Tests}
\begin{figure}
    \centering
    \includegraphics[width=1.0\textwidth]{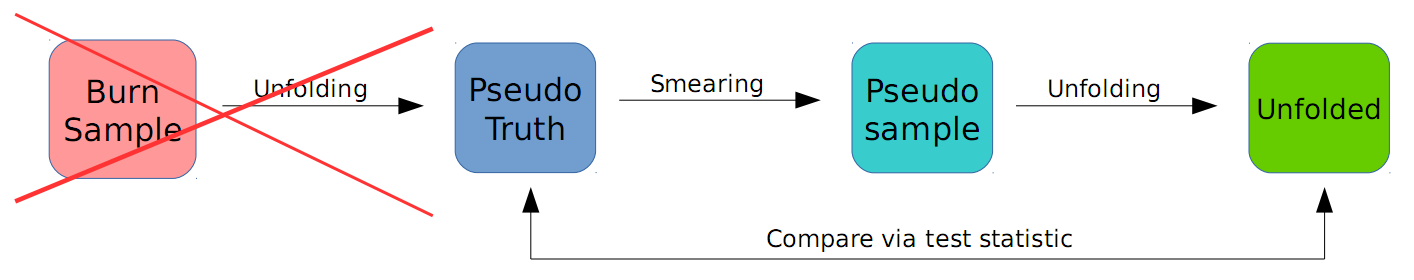}
    \caption{Schematic of the 'truth-smearing-unfolding'-test developed for this analyis. The 10\% burn sample is used as input and unfolded with the standard prescribed 4 iterations. This unfolded spectrum is then used a pseudo-truth and the response matrix is applied to it, resulting in a pseodo-measurement. This allows us to have complete knowledge of the mc truth and the reconstructed values even in the case of input data. From the pseudo measurement 200 poisson variations are drawn, each in turn compared to the unfolded now 'pseudo truth'.}
    \label{fig:tsu_schem}
\end{figure}

\begin{wraptable}{r}{8cm}
 \begin{tabular}{||c c c||} 
 \hline
  {\bf Systematic} & {\bf Value} & Prior \\ 
 \hline
 $\theta_{12}$ & $34.5^{\circ}$ & $\pm 1.1^{\circ} $\\ 
 \hline
 $\theta_{23}$ & $41^{\circ}$ & $\pm 0.11^{\circ} $\\ 
 \hline
 $\theta_{13}$ & $8.41^{\circ}$ & $\pm 0.17^{\circ} $\\ 
 \hline
 $\Delta m_{21}$ & $7.56\cdot 10^{-5}$ eV$^2$ & $\pm 0.19$ eV$^2$ \\
 \hline
 $\Delta m_{31}$ & $2.51\cdot 10^{-3}$ e$V^2 $ & $\pm 0.19$ eV$^2$ \\
 \hline
 $\delta_{cp}$ & $252^{\circ}$ & $\pm 24^{\circ}$  \\
 \hline
 Muon Scale & 1.0 & $\pm 5\%$  \\
 \hline
 Noise Scale & 1.0 & $\pm 10\%$  \\
 \hline
 DOM eff. & 1.0 & $\pm 10\%$  \\
\hline
 Hole Ice & 25 & $\pm 5\%$  \\
 \hline
 Bulk Ice Scattering & 1.0 & $\pm 10\%$  \\
 \hline
 Bulk Ice Absorbtion & 1.0 & $\pm 10\%$  \\
 \hline
 Livetime & 4.8 yr & $\pm 1\%$ \\
 \hline
 \end{tabular}
\caption{Systematic uncertainties used in this work \cite{Aartsen:2016nxy}, values
from the original online verrsion of\cite{Salas:2017} are used.}
\label{table:systematics}
\end{wraptable}
In order to test the consistency of the unfolding algorithm, several closure tests were performed. The unfolding is stable under an MC closure test, where the input measurement is generated from the standard MC simply matches the MC input in the response matrix. Varying the shape and normalization of the input MC still renders the unfolding stable, but needing more iterations before converging on the input truth. These closure tests were performed using an MC sample where both truth and reconstruction were known. The number of iterations were set to minimize the difference between the unfolded spectrum and the input MC truth.  In an actual analysis of measured data this is not possible, a novel method was employed. The diagram in figure \ref{fig:tsu_schem} illustrates the process. A 10\% sub-sample (burn sample) was unfolded with the standard 4 iterations. The unfolded spectrum is then set as a pseudo-truth and the detector response matrix is applied to it so as to generate a pseudo measurement, which is scaled to be comparable to the full experimental statistics. This pseudo measurement is then taken as input for the unfolding algorithm and can be compared to the pseudo-truth. This is done from an ensemble consisting of 200 pseudo-experiments, drawn around the pseudo-measurement. Each is individually unfolded and can be compared to the pseudo truth. Due to processing constraints we test the unfolding up to 25 iterations.

\section{Stopping Condition}
\begin{figure}
    \centering
    \includegraphics[width=0.50\textwidth]{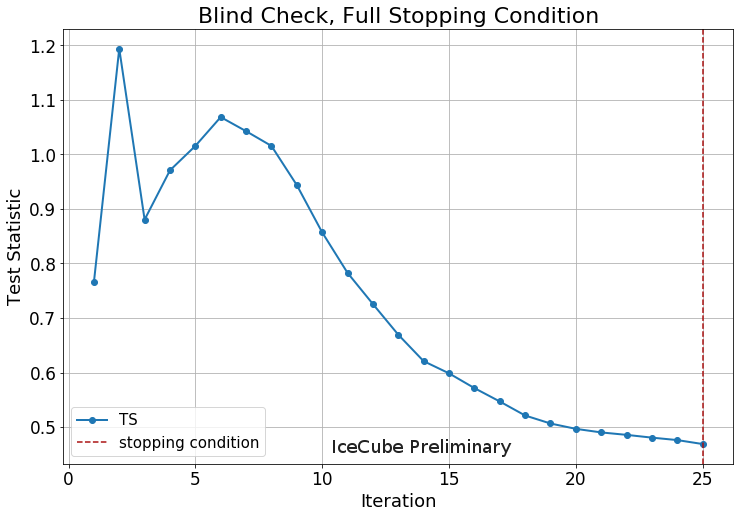}
    \caption{Stopping condition for the unfolding - the test statistic is calculated blindly including all known uncertainties, and is seen to converge with the number of iterations. The ideal stopping condition is found when the value of the test statistic is minimized. At iterations above 20 the test statistic changes only marginally, and we set the stopping condition to 25 iterations.}
    \label{fig:stop_cond}
\end{figure}

The bayesian unfolding method needs a number of iterations in order to converge on the most likely spectrum - a challenge using this method is choosing the right number of iterations to minimize both bias and statistical variance. Two methods for setting the stopping condition are calculated. The first one is based on statistical effects only, the second one includes unfolding error and systematic uncertainties. In the statistics only case, each pseudo experiment is unfolded individually and a test statistic (TS) is calculated to quantify the difference between the unfolded pseudo experiment and the pseudo truth. It is defined as follows:
\begin{equation}
    TS_{stat} = \frac{1}{NU}\sum_{i} \frac{(n_{i}U - Nu_{i})^2}{n_{i} + u_{i}}
    \label{ts_stat}
\end{equation}, 
where \(n_{i}\) and \(u_{i}\) are ther number of expected and unfolded events in bin i, respectively, while N and U represent the total number of expected and unfolded events, respectively.
This TS is calculated for each of the 200 pseudo experiments and the average is plotted as a function of iterations, as seen in fig, \ref{fig:stop_cond}. The error is taken as the square root of the variance of the 200 TS ensemble. The stopping condition is imposed when the change in TS drops to be lower than the \(1\sigma\) error bar on the TS:
\begin{equation}
    n_{stop}: \Delta TS_{stat} \leq \sigma_{n}^{TS}.
\end{equation}
At which point a distinction between the n'th and n+1'th iteration becomes difficult from a statistical point of view A clear improvement in consistency between the unfolded spectrum and the input pseudo-truth is seen with the number of iterations, as shown in fig. \ref{fig:stop_cond} .
In the full systematics case, we instead minimize the width of the total uncertainty band on the unfolding, including both the confidence percentiles from the 1400 trials as well as statistical uncertainty and the small impact from the unfolding error:
\begin{equation}
    \sigma_{tot}^{2} = \sigma_{stat}^2 + \sigma_{syst}^2 + \sigma_{bias}^2
    \label{sigma_tot}
\end{equation}This in order to overcome the potential case where fluctuations due to systematics get picked up and become enhanced by the iterative unfolding. This calculation is performed blind and the stopping condition is set before unblinding the full analysis. Since the two stopping condition distributions can exhibit different functional behaviours, we utilize the following hierarchy to set the stopping condition: \\
\indent{1) If the statistics-only TS-distribution in eq. \ref{ts_stat} shows divergent behaviour: Use 4 iterations.}\\
\indent{2) If the statistics-only TS-distribution in eq. \ref{ts_stat} shows convergent behaviour: } \\
\indent{  \indent{Minimize full uncertainty band in eq. \ref{sigma_tot} instead.}} \\
\indent{3) In case of systematics dominance after 1 iteration: Use statistics only stopping condition.} \\
\indent{4) Otherwise: Minimize TS based on the uncertainty band in eq. \ref{sigma_tot}}, as described above. \\
In our blind test with the full data sample, the statistics only distribution converges, and we set the stopping condition by minimizing the TS based on the full uncertainty band. The distribution is shown in fig. \ref{fig:stop_cond}\\

\section{Event Rate Unfolding}
\begin{figure}
    \includegraphics[width=0.50\textwidth]{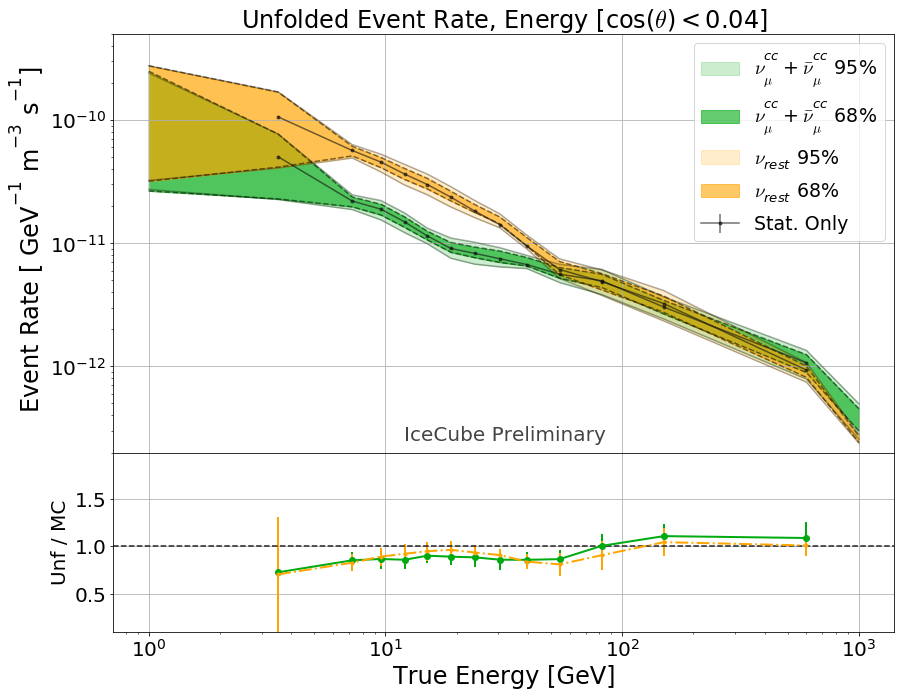}
    \includegraphics[width=0.50\textwidth]{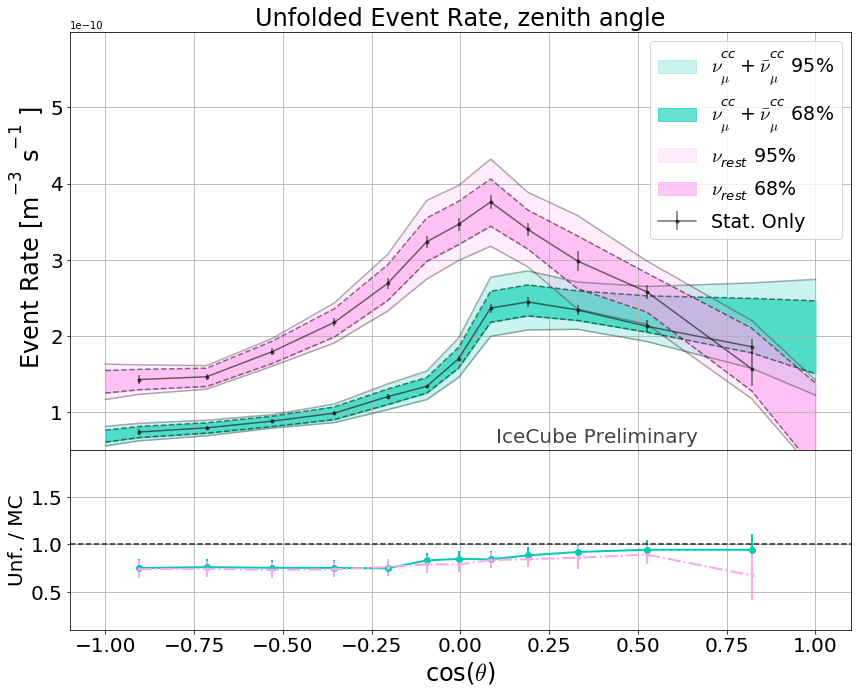}
    \caption{Unfolded spectra of in-ice event rate by volume. Left: Energy. Right: \(\cos(\theta_{z})\) Error bands show the full uncertainty including statistical, systematic and unfolding errors. The black central line show the unfolded spectrum when keeping all systematics at their nominal values, and the error bars represent the statistical uncertainty. These are too tiny to been seen in the energy spectrum due to the large range of the y-axis. The uncertainty bands have been estimated from ~1400 trial sets of the systematic parameters and show the 68\% and 95\% confidence intervals. The green and turquoise bands represents the combined \(\nu_{\mu}\) and \(\bar{\nu}_{\mu}\) charged current interactions, the orange and pink bands contains all other flavor and interaction combinations. The lower panel shows the ratio of the unfolded spectrum versus the expectation based on the Honda flux model. The spectrum is taken as the average of the unfolded systematic trial sets. The unfolded measurement generally agrees with the Honda model within the error bars in the down going region, but does show some deviation in the up going region.}
    \label{fig:unf_spec}
\end{figure}

\begin{figure}
    \includegraphics[width=0.50\textwidth]{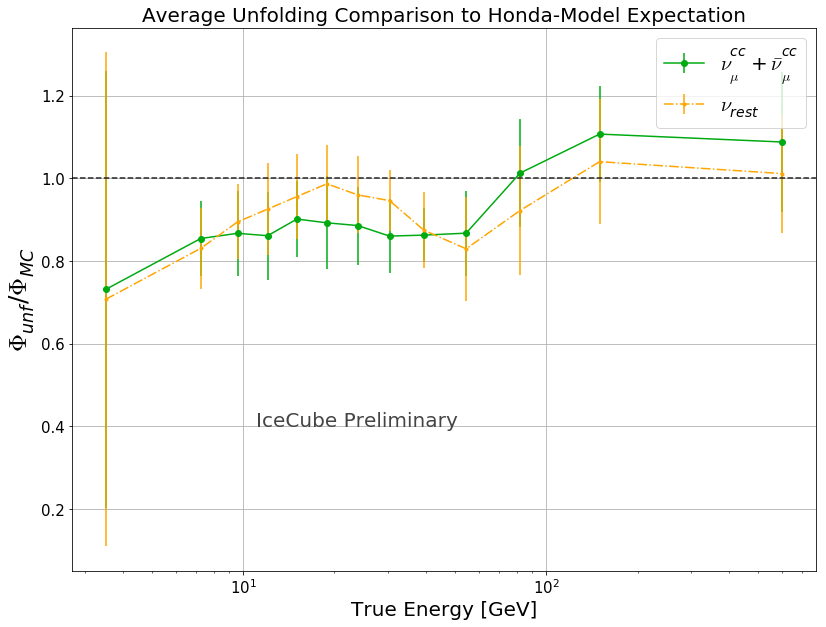}
    \includegraphics[width=0.50\textwidth]{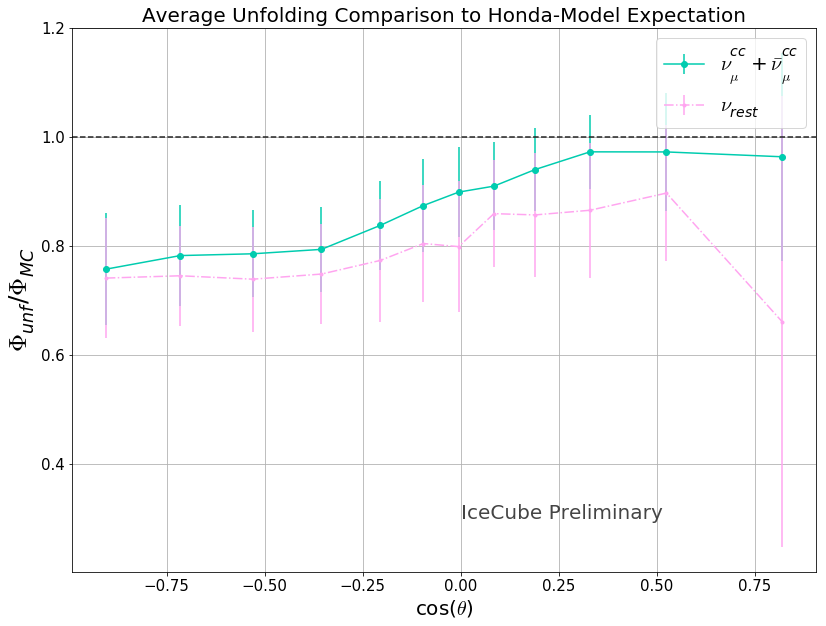}
    \caption{Unfolded flux energy spectrum versus expectation based on the Honda flux model. Left: Energy. Right: \(\cos(\theta)\). The energy spectrum is taken for the up going region up to \(\cos(\theta) < 0.04\). The spectrum is taken as the average of the unfolded systematic trial sets. The IceCube measurement mostly agrees with the MC prediction  within the error bars, but does show some deviation at lower energies and up going region.}
    \label{fig:unf_ratio}
\end{figure}

Unfolding to iteration 25 as given by the stopping condition, the estimated true unfolded event rate by volume is shown for both energy and \(\cos(\theta_{z})\) for both of our unfolded PID channels, in fig \ref{fig:unf_spec}. 
The results shown here are minimally biased as the unfolding compensates for MC bias. As such the unfolded event rate still include information on the oscillation parameters, the interaction cross sections and the ice density. The ratio of our unfolded result vs our MC event rate by volume expectation based on the Genie interaction model \cite{genie:2009} the HKKM atmospheric flux model \cite{Honda:2015}, when considering the oscillation parameters in table \ref{table:systematics}, is shown in figure \ref{fig:unf_ratio}. Due to the nature of the unfolding matrix, and the imperfect particle identification, the unfolded channels are correlated. While the energy spectrum above 10 GeV shows good agreement with the model prediction, we do find some tension at very low energies and in the up going region.

\subsection{Flux Unfolding} 
\begin{figure}
    \includegraphics[width=0.50\textwidth]{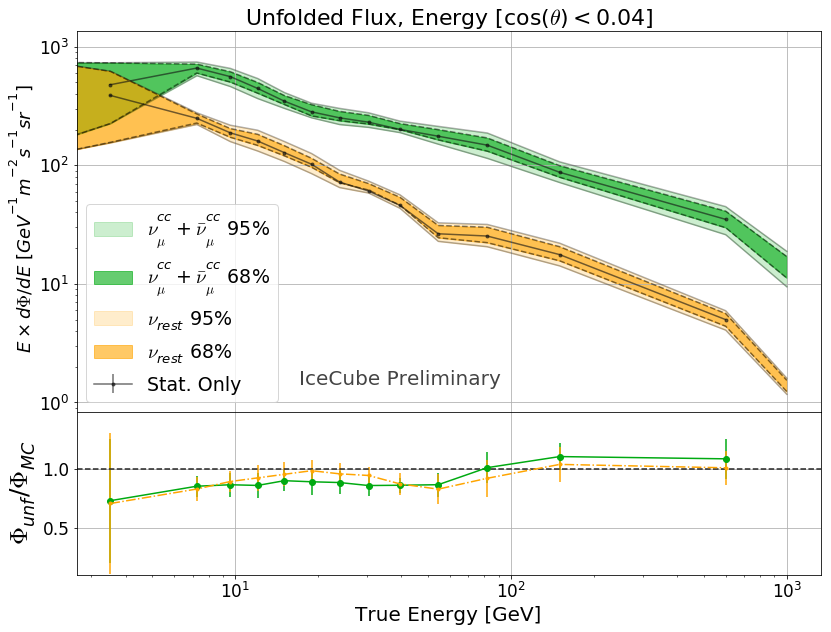}
    \includegraphics[width=0.50\textwidth]{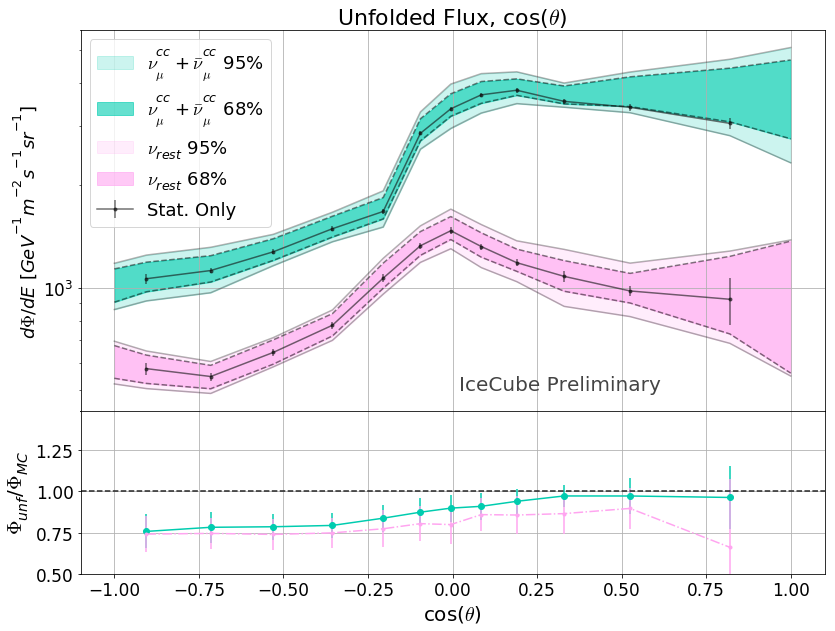}
    \caption{Unfolded flux energy spectrum. Error bands show the full uncertainty including statistical, systematic and unfolding errors. The black central line show the unfolded spectrum when keeping all systematics at their nominal values, and the error bars represent the statistical uncertainty. These are too tiny to been seen due to the large range of the y-axis. The uncertainty bands have been estimated from ~75 trial sets of the systematic parameters, due to time constraints, and show the 68\% and 95\% confidence intervals. The green and turquoise band represents the combined  \(\nu_{\mu}\) and \(\bar{\nu}_{\mu}\) charged current interactions, the orange and pink bands contains all other flavor and interaction combinations. The lower panel shows the ratio of the unfolded flux spectrum versus the expectation based on the Honda flux model. The spectrum is taken as the average of the unfolded systematic trial sets. The unfolded measurement generally agrees with the Honda model within the error bars in the down going region, but does show some deviation in the up going region.}
    \label{fig:unf_spec_flux}
\end{figure}

As an addendum to the main result, we also unfold the atmospheric neutrino flux. The unfolded quantity is a matter of the weighting of the truth side of the response matrix, and it is relatively straight forward to change the unfolded quantity. Using the same number of iterations a separate unfolding is performed, unfolding directly to the atmospheric neutrino flux instead of the volumetric event rate. The unfolded flux is shown in fig. \ref{fig:unf_spec_flux}

\section{Conclusion and Outlook}
We have presented a model agnositc measurement of the atmospheric neutrino flux at the South Pole, using IceCube/DeepCore. This measurement has the event statistics to make detailed statements of the atmospheric neutrino flux, in the region of interest to investigations of both the oscillation parameters and the interaction cross section.  Already at this early stage the results are competetive. The measurement precision is completely dominated by the systematic uncertainties related to the antarctic ice and the optical modules. These properties are under continuous evaluation by the IceCube collaboration and any improvement will also improve the results contained in this analysis. In the future this work is planned to be updated with a larger data sample, and more stringent constraints on the systematic parameters.

\bibliographystyle{ICRC}
\bibliography{references}

%

\end{document}